\documentclass [prd,twocolumn,superscriptaddress,floatfix,showpacs,amsmath,letterpaper] {revtex4}

\usepackage{graphicx}
\usepackage{dcolumn}
\usepackage{bm}
\usepackage{color}

\newcommand{\al}{$\bar{\Lambda}$}

\begin{document}

\title {Longitudinal Spin Transfer to \bm {$\Lambda$} and \bm {$\bar{\Lambda}$} Hyperons in Polarized Proton-Proton Collisions at \bm {$\sqrt{s} = 200\,\mathrm{GeV}$}}

\affiliation{Argonne National Laboratory, Argonne, Illinois 60439, USA}
\affiliation{University of Birmingham, Birmingham, United Kingdom}
\affiliation{Brookhaven National Laboratory, Upton, New York 11973, USA}
\affiliation{University of California, Berkeley, California 94720, USA}
\affiliation{University of California, Davis, California 95616, USA}
\affiliation{University of California, Los Angeles, California 90095, USA}
\affiliation{Universidade Estadual de Campinas, Sao Paulo, Brazil}
\affiliation{University of Illinois at Chicago, Chicago, Illinois 60607, USA}
\affiliation{Creighton University, Omaha, Nebraska 68178, USA}
\affiliation{Czech Technical University in Prague, FNSPE, Prague, 115 19, Czech Republic}
\affiliation{Nuclear Physics Institute AS CR, 250 68 \v{R}e\v{z}/Prague, Czech Republic}
\affiliation{University of Frankfurt, Frankfurt, Germany}
\affiliation{Institute of Physics, Bhubaneswar 751005, India}
\affiliation{Indian Institute of Technology, Mumbai, India}
\affiliation{Indiana University, Bloomington, Indiana 47408, USA}
\affiliation{University of Jammu, Jammu 180001, India}
\affiliation{Joint Institute for Nuclear Research, Dubna, 141 980, Russia}
\affiliation{Kent State University, Kent, Ohio 44242, USA}
\affiliation{University of Kentucky, Lexington, Kentucky, 40506-0055, USA}
\affiliation{Institute of Modern Physics, Lanzhou, China}
\affiliation{Lawrence Berkeley National Laboratory, Berkeley, California 94720, USA}
\affiliation{Massachusetts Institute of Technology, Cambridge, MA 02139-4307, USA}
\affiliation{Max-Planck-Institut f\"ur Physik, Munich, Germany}
\affiliation{Michigan State University, East Lansing, Michigan 48824, USA}
\affiliation{Moscow Engineering Physics Institute, Moscow Russia}
\affiliation{City College of New York, New York City, New York 10031, USA}
\affiliation{NIKHEF and Utrecht University, Amsterdam, The Netherlands}
\affiliation{Ohio State University, Columbus, Ohio 43210, USA}
\affiliation{Old Dominion University, Norfolk, VA, 23529, USA}
\affiliation{Panjab University, Chandigarh 160014, India}
\affiliation{Pennsylvania State University, University Park, Pennsylvania 16802, USA}
\affiliation{Institute of High Energy Physics, Protvino, Russia}
\affiliation{Purdue University, West Lafayette, Indiana 47907, USA}
\affiliation{Pusan National University, Pusan, Republic of Korea}
\affiliation{University of Rajasthan, Jaipur 302004, India}
\affiliation{Rice University, Houston, Texas 77251, USA}
\affiliation{Universidade de Sao Paulo, Sao Paulo, Brazil}
\affiliation{University of Science \& Technology of China, Hefei 230026, China}
\affiliation{Shandong University, Jinan, Shandong 250100, China}
\affiliation{Shanghai Institute of Applied Physics, Shanghai 201800, China}
\affiliation{SUBATECH, Nantes, France}
\affiliation{Texas A\&M University, College Station, Texas 77843, USA}
\affiliation{University of Texas, Austin, Texas 78712, USA}
\affiliation{Tsinghua University, Beijing 100084, China}
\affiliation{United States Naval Academy, Annapolis, MD 21402, USA}
\affiliation{Valparaiso University, Valparaiso, Indiana 46383, USA}
\affiliation{Variable Energy Cyclotron Centre, Kolkata 700064, India}
\affiliation{Warsaw University of Technology, Warsaw, Poland}
\affiliation{University of Washington, Seattle, Washington 98195, USA}
\affiliation{Wayne State University, Detroit, Michigan 48201, USA}
\affiliation{Institute of Particle Physics, CCNU (HZNU), Wuhan 430079, China}
\affiliation{Yale University, New Haven, Connecticut 06520, USA}
\affiliation{University of Zagreb, Zagreb, HR-10002, Croatia}

\author{B.~I.~Abelev}\affiliation{University of Illinois at Chicago, Chicago, Illinois 60607, USA}
\author{M.~M.~Aggarwal}\affiliation{Panjab University, Chandigarh 160014, India}
\author{Z.~Ahammed}\affiliation{Variable Energy Cyclotron Centre, Kolkata 700064, India}
\author{A.~V.~Alakhverdyants}\affiliation{Joint Institute for Nuclear Research, Dubna, 141 980, Russia}
\author{B.~D.~Anderson}\affiliation{Kent State University, Kent, Ohio 44242, USA}
\author{D.~Arkhipkin}\affiliation{Brookhaven National Laboratory, Upton, New York 11973, USA}
\author{G.~S.~Averichev}\affiliation{Joint Institute for Nuclear Research, Dubna, 141 980, Russia}
\author{J.~Balewski}\affiliation{Massachusetts Institute of Technology, Cambridge, MA 02139-4307, USA}
\author{O.~Barannikova}\affiliation{University of Illinois at Chicago, Chicago, Illinois 60607, USA}
\author{L.~S.~Barnby}\affiliation{University of Birmingham, Birmingham, United Kingdom}
% ...

\collaboration{STAR Collaboration}\noaffiliation


\vspace*{0.3cm}

\begin{abstract}
The longitudinal spin transfer, $D_{LL}$, from high energy polarized protons to $\Lambda$ and $\bar{\Lambda}$ hyperons has been measured for the first time in proton-proton collisions at $\sqrt{s} = 200\,\mathrm{GeV}$ with the STAR detector at RHIC.
The measurements cover pseudorapidity, $\eta$, in the range $\left|\eta\right| < 1.2$ and transverse momenta, $p_\mathrm{T}$, up to $4\,\mathrm{GeV}/c$.
The longitudinal spin transfer is found to be $D_{LL}= -0.03\pm 0.13(\mathrm{stat}) \pm 0.04(\mathrm{syst})$
 for inclusive $\Lambda$ and $D_{LL} = -0.12 \pm 0.08(\mathrm{stat}) \pm 0.03(\mathrm{syst})$ for inclusive $\bar{\Lambda}$ hyperons with  $\left<\eta\right> = 0.5$ and $\left<p_\mathrm{T}\right> = 3.7\,\mathrm{GeV}/c$.
The dependence on $\eta$ and $p_\mathrm{T}$ is presented.

\end{abstract}
\maketitle


The longitudinal spin transfer to $\Lambda$ and $\bar{\Lambda}$ hyperons has been studied in $e^+e^-$ collisions at LEP~\cite{aleph,opal} and in deep-inelastic scattering of neutrinos~\cite{nomad}, polarized muons~\cite{e665,compass}, and polarized positrons~\cite{hermes} on unpolarized targets.
The phenomenon is understood to originate from different physical mechanisms in the different types of reactions.
At LEP, the fragmentation of highly-polarized strange quark and anti-quark pairs is expected to dominate.
In deep-inelastic scattering, the spin transfer from struck quarks and target fragments is expected to play an important role~\cite{phenomenology}.

In this paper, we study the longitudinal spin transfer, $D_{LL}$, to $\Lambda$ and $\bar{\Lambda}$ hyperons produced in polarized proton-proton collisions at $\sqrt{s} = 200\,\mathrm{GeV}$ center-of-mass energy,
\begin{equation}
D_{LL}\equiv \frac
{\sigma_{p^+p \to  \Lambda ^+ X}-\sigma_{p^+p \to  \Lambda ^-X}}
{\sigma_{p^+p \to  \Lambda ^+ X}+\sigma_{p^+p \to  \Lambda ^-X}},
\label{gener1}
\end{equation}
where the superscripts $+$ and $-$ denote helicity.
The production cross section has been measured for transverse momenta, $p_\mathrm{T}$, up to about $5\,\mathrm{GeV}/c$ and is 
reasonably well described by perturbative QCD evaluations for a suitable choice of fragmentation functions~\cite{Abelev:2006cs}.
Within this framework, the production cross sections are described in terms of calculable partonic cross sections and non-perturbative parton distribution and fragmentation functions.
The spin transfer $D_{LL}$ is thus expected to be sensitive to polarized parton distribution functions and polarized fragmentation functions.
Present data are too scarce to adequately constrain the polarized fragmentation functions. 
Sizable uncertainties also remain in the polarized parton distributions, particularly in the polarized sea quark and gluon distributions.
This is reflected in the model predictions for $D_{LL}$ at RHIC~\cite{deFlorian:1998ba,Boros:2000ex,Ma:2001na,Xu:2002hz,Xu:2005ru}.

The polarization of  $\Lambda$ $(\bar{\Lambda})$ hyperons, ${P}_{\Lambda(\bar\Lambda)}$, can be measured via the weak decay channel $\Lambda \to p \pi^-$ $(\bar \Lambda \to \bar p \pi^+)$ from the angular distribution of the final state,
\begin{equation}
\frac{\mathrm{d}N}{\mathrm{d} \cos{\theta}^*}=\frac{
\sigma\,\mathcal{L}\,A}{2}
(1+\alpha_{\Lambda(\bar\Lambda)} P_{\Lambda(\bar\Lambda)}
\cos{\theta}^*),
\label{ideal}
\end{equation}
where 
$\sigma$ is the (differential) production cross section, $\mathcal{L}$ is the integrated luminosity, $A$ is the detector acceptance, which may vary with $\theta^*$ as well as other observables, and $\alpha_{\Lambda}$=$-\alpha_{\bar{\Lambda}} = 0.642 \pm 0.013$~\cite{Amsler:2008zz} is the weak decay parameter. 
In this paper, the polarization along the $\Lambda$ $(\bar\Lambda)$ momentum direction is considered and $\theta^*$ is the angle between the polarization direction and the \mbox{(anti-)proton} momentum in the $\Lambda$ ($\bar\Lambda$) rest frame.
The spin transfer $D_{LL}$ in Eq.~(\ref{gener1}) is identical to ${P}_{\Lambda(\bar\Lambda)}$ if the proton beam polarization is maximal.


The data were collected at the Relativistic Heavy Ion  Collider (RHIC) with the Solenoidal Tracker at RHIC (STAR)~\cite{Ackermann:2002ad} in the year 2005.
An integrated luminosity of 2\,$\mathrm{pb}^{-1}$ was sampled with  longitudinal proton beam spin configurations.
The degree of proton polarization was measured for each beam and each beam fill using Coulomb-Nuclear Interference (CNI) proton-carbon polarimeters~\cite{Jinnouchi:2004up}, which were calibrated \emph{in situ} using a polarized atomic hydrogen gas-jet target~\cite{Okada:2006dd}.
The average longitudinal polarizations for the two beams were $52 \pm 3\%$ and $48 \pm 3\%$ for the analyzed data.
Different beam spin configurations were used for successive beam bunches and the pattern was changed between beam fills to minimize systematic uncertainties.
The data were sorted by beam spin configuration.

Beam-Beam Counters (BBC) at both sides of the STAR interaction region were used to signal proton beam collision events, to measure the relative  luminosities for the different beam spin configurations, and to  determine the size of any residual transverse beam polarization components at the STAR interaction region~\cite{Kiryluk:2005gg}.
The BBC proton collision signal defined the minimum bias (MB) trigger condition.
The data presented here were recorded with the MB trigger condition and with two additional trigger conditions.
A high-tower (HT) trigger condition required the BBC proton collision signal in coincidence with  a transverse energy deposit $E_\mathrm{T} > 2.6\,\mathrm{GeV}$ in at least one Barrel Electromagnetic Calorimeter (BEMC)~\cite{Beddo:2002zx} tower, covering $\Delta\eta \times \Delta\phi=0.05 \times 0.05$ in pseudorapidity, $\eta$, and azimuthal angle, $\phi$.
A jet-patch (JP) trigger condition imposed the MB condition in coincidence with an energy deposit $E_\mathrm{T} > 6.5\, \mathrm{GeV}$ in at least one of six BEMC patches each covering $\Delta\eta \times \Delta\phi=1 \times 1$.
The total BEMC coverage was $0 < \eta < 1$ and $0 < \phi < 2\pi$ in 2005.
Charged particle tracks in the 0.5\,T magnetic field were measured with the Time Projection Chamber (TPC)~\cite{Anderson:2003ur}, covering $0 < \phi < 2\pi$ and $|\eta|\lesssim 1.3$.
The measurement of specific  energy loss, $\mathrm{d}E/\mathrm{d}x$, in the TPC gas provided particle identification~\cite{Shao:2005iu}.

The  $\Lambda$ and $\bar{\Lambda}$ candidates were identified from the topology of their dominant weak decay channels, $\Lambda \to p \pi^-$ and $\bar \Lambda \to \bar p \pi^+$, each having a branching ratio of 63.9\%~\cite{Amsler:2008zz}. 
The procedure closely resembled the one used in the cross section measurement reported in Ref.~\cite{Abelev:2006cs}.
The reconstructed event vertex was required to be along the beam axis and within 60\,cm of the TPC center to ensure uniform tracking efficiency.
A sample of $1.8 \times 10^6$ events satisfying the MB trigger condition was analyzed.
In addition, $2.5 \times 10^6$ events were analyzed satisfying the HT trigger condition and $3.2 \times 10^6$ events satisfying the JP trigger condition.
About $0.3 \times 10^6$ of these events satisfied both HT and JP trigger requirements.
A search was made in each event to find (anti-)proton and pion tracks of opposite curvature.
The tracks were then paired to form a $\Lambda (\bar{\Lambda})$ candidate and topological selections were applied to reduce background.
The selections included criteria for the distance of closest approach between the paired tracks and the distance between the point of closest approach and the beam collision vertex, and demanded that the momentum sum of the track pair pointed at the collision vertex.
The criteria were tuned to preserve the signal while reducing the background fraction to 10\% or less.

\begin{figure}
\begin{center}
 \includegraphics[height=.18\textheight]{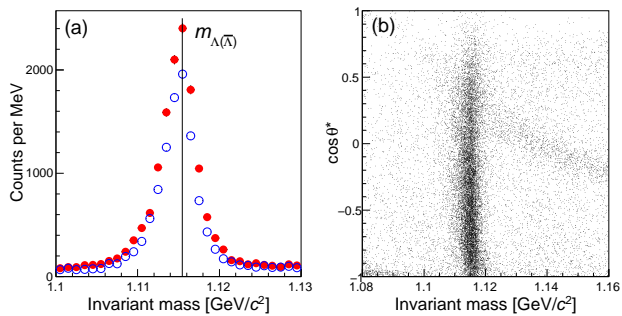}
 \caption{(Color online)(a) The invariant mass distribution of $\Lambda$ (filled circles) and $\bar \Lambda$ (open circles) candidates from reconstructed $p + \pi^-$ and $\bar p + \pi^+$ track pairs in 2005 MB data after topological selections.
(b) The invariant mass distribution versus $\cos\theta^*$ for $\Lambda$.}
\label{mass}
\end{center}
\end{figure}

Figure~\ref{mass}(a) shows the invariant mass distribution for the $\Lambda$ (filled circles) and $\bar\Lambda$ (open circles) candidates reconstructed from MB data with $|\eta| < 1.2$ and $0.3 < p_\mathrm{T} < 3\,\mathrm{GeV}/c$.
The mean values of the $\Lambda$ and $\bar{\Lambda}$ mass distributions are in agreement with the PDG mass value $m_{\Lambda(\bar\Lambda)} = 1.11568\,\mathrm{GeV}/c^2$~\cite{Amsler:2008zz}.
Figure~\ref{mass}(b) shows the same invariant mass distribution versus $\cos\theta^*$ for the $\Lambda$ candidates.
The number of $\Lambda$ candidates varies with $\cos\theta^*$ because of detector acceptance.
The small variation of the reconstructed invariant mass with $\cos\theta^*$ is understood to originate from detector resolution.
In addition to signal and combinatorial background, backgrounds are seen of misidentified $e^+e^-$ pairs at low invariant mass values near $\cos\theta^*=-1.0$ and of misidentified $K_{S}^0$ in a diagonal band at high invariant mass values and $\cos\theta^*>-0.2$.
About $1.2 \times 10^4\ \Lambda$ and $1.0 \times 10^4\ \bar{\Lambda}$ candidates with reconstructed invariant mass $1.109< m <1.121\,\mathrm{GeV}/c^2$ were kept for further analysis.
The average residual background fraction was determined to be about 9\% by averaging the candidate counts in the mass intervals $1.094 < m < 1.103\,\mathrm{GeV}/c^2$ and $1.127 < m < 1.136\,\mathrm{GeV}/c^2$.

The observed spectra are affected by detector resolution and acceptance.
To minimize the uncertainty associated with acceptance effects, $D_{LL}$ has been extracted in small intervals in $\cos\theta^*$ from the ratio:
\begin{equation}
D_{LL}=\frac{1}{\alpha P_\mathrm{beam} \left< \cos \theta^* \right>} \frac{N^+ -R N^-} {N^+ + R N^-},
\label{eq_dll}
\end{equation}
where $\alpha$ is the decay parameter, $P_\mathrm{beam}$ is the measured polarization for either RHIC beam, and $\left< \cos \theta^* \right>$ denotes the average in the $\cos\theta^*$ interval.
Eq.~(\ref{eq_dll}) follows from Eqs.~(\ref{gener1}) and ~(\ref{ideal}), and parity conservation in the hyperon production processes considered here.
The single spin hyperon yield $N^+$ was obtained by summing the double spin yields $n^{++}$ and $n^{+-}$ 
weighted by the relative luminosity for the ${++}$ and ${+-}$ beam spin configurations.
The yield $N^-$ was obtained
in a similar way
from $n^{-+}$ and $n^{--}$, and $R$ denotes the relative luminosity ratio 
to normalize $N^+$ and $N^-$.
The single spin yield $N^+$ can also be determined from the alternative combination of double spin yields, $n^{++}$ and $n^{-+}$, as if the other beam is (un-)polarized.
In this case $N^-$ is obtained from $n^{+-}$ and $n^{--}$.

The yields $N^+$ and $N^-$ were determined for each $\cos\theta^*$ interval from the observed $\Lambda$ and $\bar{\Lambda}$ candidate yields in the mass interval from 1.109 to 1.121\,GeV/c$^2$. 
The corresponding raw values $D^\mathrm{raw}_{LL}$ were averaged over the entire $\cos\theta^*$ range.
The obtained $D^\mathrm{raw}_{LL}$ values and their statistical uncertainties were then corrected for (unpolarized) background dilution according to $D_{LL} = D^\mathrm{raw}_{LL}/(1-r)$, where $r$ is the average background fraction.
No significant spin transfer asymmetry was observed for the yields in the sideband mass intervals $1.094 < m < 1.103\,\mathrm{GeV}/c^2$ and $1.127 < m < 1.136\,\mathrm{GeV}/c^2$, and thus no further correction was applied to $D_{LL}$.
However, a contribution was included in the estimated systematic uncertainty of the $D_{LL}$ measurement to account for the possibility that the background could nevertheless be polarized.
Its size was determined by the precision of the spin transfer asymmetries for the sideband mass intervals.
The $D_{LL}$ results obtained with either of the beams polarized were combined, taking into account the overlap in the event samples.

\begin{figure}
\begin{center}
  \includegraphics[height=.50\textheight]{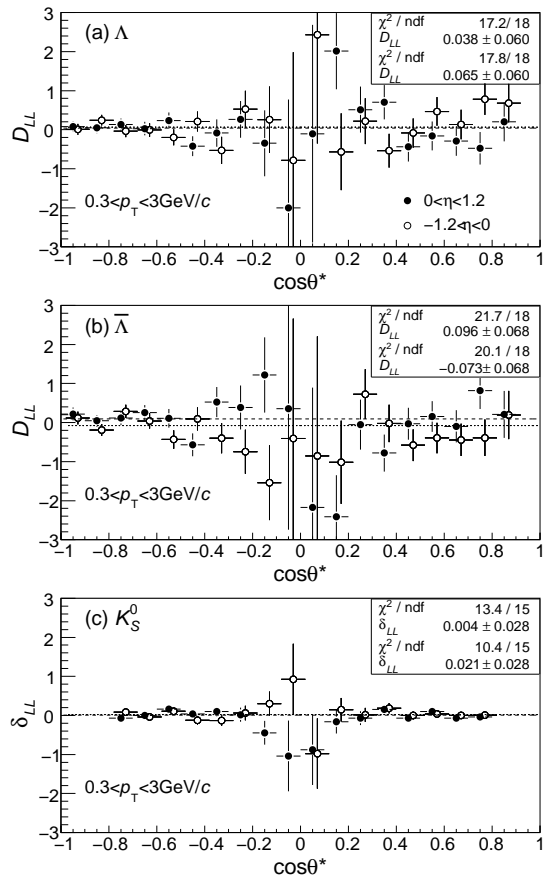}
\caption{(Color online) The spin transfer $D_{LL}$ versus $\cos\theta^*$ for (a) $\Lambda$ and (b) $\bar \Lambda$ hyperons, and (c) the spin asymmetry $\delta_{LL}$ for the control sample of $K^0_S$ mesons versus $\cos\theta^*$. The filled circles show the results for positive pseudorapidities $\eta$ with respect to the polarized beam and the open circles show the results for negative $\eta$. 
Only statistical uncertainties are shown.
The data points with negative $\eta$ have been shifted slightly in $\cos\theta^*$ for clarity.
The indicated values of $\chi^2$ and the spin transfer are for the data with positive and negative $\eta$, respectively.
}
\label{cosin_fit}
\end{center}
\end{figure}

Figure \ref{cosin_fit}(a) shows the combined $D_{LL}$ results from the MB data sample versus $\cos\theta^*$ for the inclusive production of $\Lambda$ hyperons with $0.3 < p_\mathrm{T} < 3\,\mathrm{GeV}/c$ and $0 < \eta < 1.2$ and $ -1.2 < \eta < 0$.
The results for the \al\ hyperon are shown in Fig. \ref{cosin_fit}(b).
Positive $\eta$ is defined along the direction of the incident polarized beam.
Fewer than 50 counts were observed for $\cos\theta^* > 0.9$ and this interval was discarded for this reason.
The extracted $D_{LL}$ is constant with $\cos\theta^*$, as expected and confirmed by the quality of fit.
In addition, a null-measurement was performed of the spin transfer for the spinless $K_S^0$ meson, which has a similar event topology.
The $K_S^0$ candidate yields for $|\cos\theta^*| > 0.8$ were discarded since they have sizable $\Lambda (\bar{\Lambda}$) backgrounds.
The result, $\delta_{LL}$, obtained with an artificial weak decay parameter $\alpha_{K_S^0} = 1$, was found consistent with no spin transfer, as shown in Fig.~\ref{cosin_fit}(c).
The analysis was furthermore tested with simulated $\Lambda$ data having a non-zero $D_{LL}$
and the $D_{LL}$ input to the simulation was extracted successfully.

\begin{figure}
\begin{center}
  \includegraphics[height=.32\textheight]{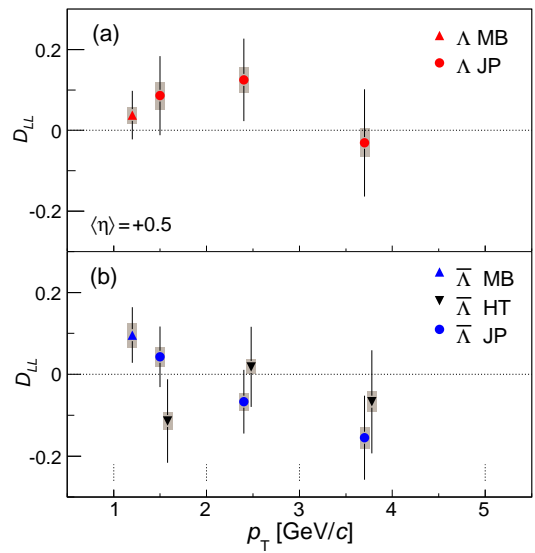}
\caption{
(Color online) The spin transfer $D_{LL}$ to (a) $\Lambda$ and (b) $\bar \Lambda$  
hyperons produced at positive pseudorapidity with respect to the  
polarized proton beam from MB, JP, and HT data versus hyperon  
transverse momenta $p_\mathrm{T}$.  The sizes of the statistical and systematic  
uncertainties are indicated by the vertical bars and bands, respectively.
For clarity, the HT data points have been shifted slightly in $p_\mathrm{T}$.
The dotted vertical lines indicate the $p_\mathrm{T}$ intervals in the analysis of HT and JP data.
}
\label{DLL_3trg}
\end{center}
\end{figure}

In addition to the MB data, HT and JP data were analyzed.
These data were recorded with trigger conditions that required large energy deposits in the BEMC and thus preferentially selected events with a hard collision.
The trigger conditions did not require a highly energetic $\Lambda$ or $\bar{\Lambda}$.
The HT and JP data samples may thus be biased.
To minimize the effects of trigger bias, the HT event sample was restricted to $\Lambda$ or $\bar{\Lambda}$ candidates whose decay (anti-)proton track intersected a BEMC tower that fulfilled the trigger condition.
About 50\% of the $\bar{\Lambda}$ and only 3\% of the $\Lambda$ candidate events in the analysis pass this selection.
This is qualitatively consistent with the annihilation of anti-protons in the BEMC.
The $\bar\Lambda$ sample that was selected in this way thus directly triggered the experiment read-out.  
It contains about 1.0$\times 10^4$ \al\ candidates with $1 < p_\mathrm{T} < 5\,\mathrm{GeV}/c$ and a residual background of about 5\%.

In the case of the JP triggered sample, events were selected with at least one reconstructed jet that pointed to a triggered jet patch.
The same jet reconstruction was used as in Refs.~\cite{Abelev:2006uq,Abelev:2007vt}.
Jets outside the BEMC acceptance were rejected.
The $\Lambda$ and $\bar{\Lambda}$ candidates  whose reconstructed $\eta$ and $\phi$ fell within the jet cone of radius $r_\mathrm{cone} = \sqrt{(\Delta\eta)^2 + (\Delta\phi)^2} = 0.4$ were retained for further analysis.
In most cases, the decay (anti-)proton track is part of the reconstructed jet, whereas the decay pion track is not.
No correction was made to the jet finding and reconstruction for this effect.
About $1.3\times10^4~\Lambda$ and $2.1\times10^4~\bar{\Lambda}$ candidates with $1 < p_\mathrm{T} < 5\,\mathrm{GeV}/c$ remain after selections.
The residual background is estimated to be 13\% for $\Lambda$ and 9\% for $\bar{\Lambda}$ candidates.

The $D_{LL}$ analyses of the HT and JP samples are identical to that of the MB sample, and the resulting $D_{LL}$ averaged over $\cos\theta^*$ have similar fit quality.
The analysis of the corresponding $K_S^0$ samples shows no evidence for unaccounted systematics.
The comparison of $D_{LL}$ from MB, HT, and JP data versus $p_\mathrm{T}$ for positive $\eta$ is shown in Fig.~\ref{DLL_3trg}.
Feynman $x$ is on average about $0.02$ in the interval at highest $p_\mathrm{T}$.

The contributions from the uncertainties in decay parameter $\alpha$ and in the measurements of the proton beam polarization and relative luminosity ratios, as well as uncertainty caused by the aforementioned backgrounds, overlapping events (pile-up), and, in the case of the JP sample, trigger bias, were combined in quadrature to estimate the size of the total systematic uncertainties.
The effect of $\Lambda$ and $\bar{\Lambda}$ spin precession in the STAR magnetic field is negligible.
The above contributions are considered to be independent and their sizes have been estimated as described below.

The uncertainty in $\alpha_{\Lambda} = 0.642 \pm 0.013$~\cite{Amsler:2008zz} corresponds to a 2\% scale uncertainty in $D_{LL}$.
Uncertainty in the RHIC beam polarization measurements and in the polarization angles at the STAR interaction region contribute an additional 6\% scale uncertainty in $D_{LL}$.
Uncertainties in the measurement of $R$ are estimated to offset $D_{LL}$ at the level of 0.01.
Each of these uncertainties is common to the data from all trigger conditions.
The residual backgrounds in the candidate yields differ for different trigger conditions.  
As described before, $D_{LL}$ and its statistical uncertainty were corrected for unpolarized dilution and a systematic uncertainty was assigned based on the possible size of residual polarized background.  
This contribution to the systematic uncertainty in $D_{LL}$ ranges from 0.01 for the MB sample to 0.03 for highest $p_\mathrm{T}$ in the triggered sample.
The TPC data for each collision event may contain track information from multiple RHIC beam crossings.
This pile-up was studied by examining the observed signal candidate yields for different instantaneous beam luminosities and by extrapolating these yields to vanishingly small collision rates, for which pile-up is negligible.
In this way, a  possible dilution of 23\% in $D_{LL}$ was estimated for MB triggered data.
This was 5\% for the JP data and a negligible dilution was found for the HT triggered data.
The JP trigger condition biases the recorded $\Lambda$ and $\bar \Lambda$ samples.
Such effects were studied by Monte Carlo simulation of events  generated with PYTHIA 6.4~\cite{Sjostrand:2006za} and the STAR detector response package based on GEANT 3~\cite{Geant}.
To within the $\approx\!5\%$ statistical uncertainty of the simulation, no evidence was found that the JP trigger biases the $gg$, $qg$, and $qq$ scattering contributions to the yields, or that the JP trigger biases quark over gluon fragmentation.
A statistically significant reduction of about 25\% in fragmentation $z$ was observed in the simulated data when the JP trigger condition is imposed.
The corresponding bias in $D_{LL}$ is estimated to be no larger than 0.01 using $D_{LL}$ expectations from a range of models.
The simulated $z$ value increases with increasing $p_\mathrm{T}$ and $z \approx\!0.5$ for the data at highest $p_\mathrm{T}$.
The total systematic uncertainty in $D_{LL}$ is found to increase from 0.02 to 0.04 with increasing $p_\mathrm{T}$ and is smaller than the statistical uncertainty, ranging from 0.06 to 0.14, for each of the data points.

\begin{figure}[ht]
\begin{center}
\includegraphics[height=.38\textheight]{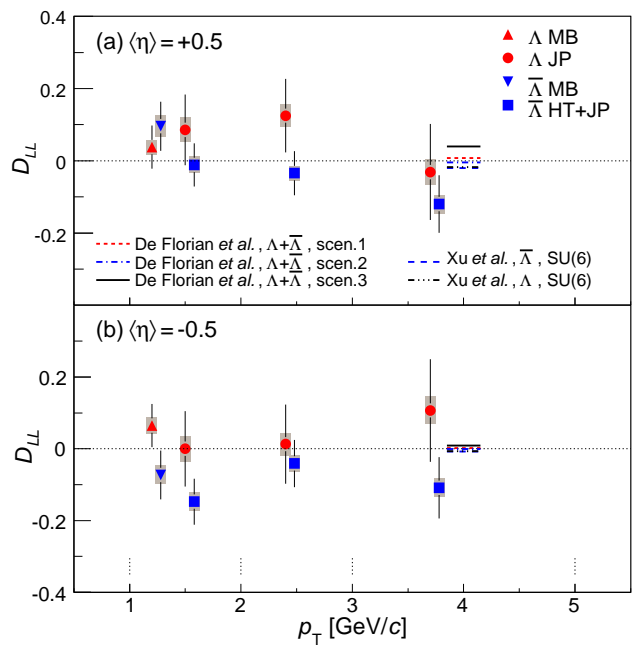}
\caption{
(Color online) Comparison of $\Lambda$ and $\bar \Lambda$ spin transfer $D_{LL}$ in  
polarized proton-proton collisions at $\sqrt{s} = 200$ GeV for (a) positive and  
(b) negative $\eta$ versus $p_\mathrm{T}$.  The vertical bars and bands indicate the  
sizes of the statistical and systematic uncertainties, respectively.
The $\bar\Lambda$ data points have been shifted slightly in $p_\mathrm{T}$ for clarity.
The dotted vertical lines indicate the $p_\mathrm{T}$ intervals in the analysis of HT and JP data.
The horizontal lines show model predictions evaluated at $\eta$ and largest $p_\mathrm{T}$ of the data.
}
\label{DLL_05eta}
\end{center}
\end{figure}

Figure~\ref{DLL_05eta} compares $\Lambda$ and $\bar{\Lambda}$ $D_{LL}$ versus $p_\mathrm{T}$ for positive and negative $\eta$.
The $\bar{\Lambda}$ results from HT and JP data have been combined.
No corrrections have been applied for possible decay contributions from heavier baryonic states.
The size of the statistical and systematic uncertainties are shown as vertical bars and shaded bands, respectively.
The $\Lambda$ and $\bar{\Lambda}$ results for $D_{LL}$ are consistent with each other and consistent with no spin transfer from the polarized proton beam to the produced $\Lambda$ and $\bar{\Lambda}$ to within the present uncertainties.
The data have $p_\mathrm{T}$ up to $4\,\mathrm{GeV}/c$, where $D_{LL}= -0.03\pm 0.13(\mathrm{stat}) \pm 0.04(\mathrm{syst})$ for the $\Lambda$ and $D_{LL} = -0.12 \pm 0.08(\mathrm{stat}) \pm 0.03(\mathrm{syst})$ for the $\bar{\Lambda}$ at $\left<\eta\right> = 0.5$.
For reference, the model predictions of Refs.~\cite{deFlorian:1998ba,Xu:2002hz,Xu:2005ru}, evaluated at $\eta=\pm 0.5$ and $p_\mathrm{T}=4\,\mathrm{GeV}/c$, are shown as horizontal lines.
The expectations of Ref.~\cite{deFlorian:1998ba} hold for $\Lambda$ and $\bar{\Lambda}$ combined and examine different polarized fragmentation scenarios, in which the strange (anti-)quark carries all or only part of the $\Lambda$ $(\bar{\Lambda})$ spin.
The model in Refs.~\cite{Xu:2002hz,Xu:2005ru} separates $\Lambda$ from $\bar{\Lambda}$ and otherwise distinguishes the direct production of the $\Lambda$ and $\bar{\Lambda}$ from the (anti-)quark in the hard scattering and the indirect production via decay of heavier (anti-)hyperons.
Both sets of expectations assume that the contribution from intrinsic gluon polarization can be neglected.
The evaluations are consistent with the present data and span a range of values that, for positive $\eta$, is similar to the experimental uncertainties.
The measurements for negative $\eta$ are less sensitive.
Since the experimental uncertainties are statistics limited, future data may be anticipated to distinguish between several of these models.

In summary, we have determined the longitudinal spin transfer to $\Lambda$ and $\bar{\Lambda}$ hyperons in $\sqrt{s} = 200\,\mathrm{GeV}$ polarized proton-proton collisions for hyperon $p_\mathrm{T}$ up to $4\,\mathrm{GeV}/c$, where earlier cross section measurements are adequately described by pQCD evaluation, and have studied the $\eta$ dependence.
The spin transfer is found to be $D_{LL}= -0.03\pm 0.13(\mathrm{stat}) \pm 0.04(\mathrm{syst})$ for $\Lambda$ and $D_{LL} = -0.12 \pm 0.08(\mathrm{stat}) \pm 0.03(\mathrm{syst})$ for $\bar{\Lambda}$ hyperons with $\left<\eta\right> = 0.5$ and $\left<p_\mathrm{T}\right> = 3.7\,\mathrm{GeV}/c$.
The longitudinal spin transfer is sensitive to the polarized parton distribution and polarized fragmentation functions.
The data correspond to an integrated luminosity of $2\,\mathrm{pb}^{-1}$ with $\approx\!50\%$ beam polarization and are limited by statistics.
The present results for $\Lambda$ and $\bar{\Lambda}$ do not provide conclusive evidence for a spin transfer signal and have uncertainties that are comparable to the variation between model expectations for the longitudinal spin transfer at RHIC.

We thank the RHIC Operations Group and RCF at BNL, the NERSC Center at LBNL and the Open Science Grid consortium for providing resources and support. This work was supported in part by the Offices of NP and HEP within the U.S. DOE Office of Science, the U.S. NSF, the Sloan Foundation, the DFG cluster of excellence `Origin and Structure of the Universe', CNRS/IN2P3, STFC and EPSRC of the United Kingdom, FAPESP CNPq of Brazil, Ministry of Ed. and Sci. of the Russian Federation, NNSFC, CAS, MoST, and MoE of China, GA and MSMT of the Czech Republic, FOM and NOW of the Netherlands, DAE, DST, and CSIR of India, Polish Ministry of Sci. and Higher Ed., Korea Research Foundation, Ministry of Sci., Ed. and Sports of the Rep. Of Croatia, Russian Ministry of Sci. and Tech, and RosAtom of Russia.

\end{document}